\documentclass[pra,aps,showpacs,twocolumn]{revtex4}
\usepackage[dvips]{graphicx}
\usepackage{latexsym}
\usepackage{amsmath}
\usepackage{hyperref}

\begin{document}

\title{Necessary and sufficient conditions of freezing phenomena
of quantum discord under phase damping}

\author{Bo You}

\affiliation{Department of Physics, Sichuan University, Chengdu 610064,
China.}

\author{Li-Xiang Cen}
\email{lixiangcen@scu.edu.cn}
\affiliation{Department of Physics, Sichuan University, Chengdu 610064,
China.}

\date{\today}

\begin{abstract}
We investigate the freezing phenomenon of quantum discord occurring
in phase damping noise processes.
By relating the expression of the time variation of
the discord to the convex function of relative entropy,
we obtain the necessary and sufficient conditions of
the phenomenon for standard Bell-diagonal states.
These conditions are applicable also to
the phenomenon occurring in a non-Markovian dephasing process. Moreover,
we show that the same condition and phenomenon coincide in a new sort of
Bell-diagonal states beyond the standard form.
\end{abstract}


\maketitle

Quantum discord is regarded as a significant characteristic of quantumness
\cite{Zurek,Vedral} and has attracted much attention in recent
literatures \cite{review}. Different from
entanglement, quantum discord could capture nonclassical correlations
arising in separable states, which has been found to exist in a quantum
algorithm without entanglement \cite{Datta,Lanyon}. Besides the fundamental
theoretic interest upon quantum mechanics itself, an appeal of the study on
this subject thereby lies in the possibility to process quantum information
in a noisy environment since pure entangled states are rather fragile in
the real world. Recent studies on the dynamics of quantum discord in various
noisy environment have manifested many fascinating features of it
\cite{Werlang,feng,Maziero,Guo,Celeri,Mazzola,MAZZOLA2,experi}. It was
shown that discord is more robust than entanglement for both Markovian and
non-Markovian dissipative processes \cite{Werlang,feng}. Moreover, the
discord of a sort of Bell-diagonal states undergoing a phase damping process
was shown to exhibit a freezing phenomenon \cite
{Mazzola,MAZZOLA2,experi}, i.e., the amount of discord
would be unaffected for a finite period with the system suffering
subsequently a sudden change from classical to quantum decoherence.

A bipartite state $\rho _{AB}$ is said to have nonzero quantum discord if $
\rho _{AB}$ can not be written as a classical-quantum form, i.e., $\rho
_{AB}\neq \sum_ip_i|i_A\rangle \langle i_A|\otimes \rho _i^B$. Based on this
benchmark, several measures have been proposed to quantify the amount of
quantum discord. An information-theoretic measure of it \cite
{Zurek,Vedral} is defined as $Q_{AB}\equiv I(\rho_{AB})-I_c(\rho _{AB})$,
where $I(\rho _{AB})=S(\rho _A)+S(\rho _B)-S(\rho _{AB})$ represents the
total correlations of $\rho _{AB}$ and $S(\rho )=-
\mathrm{{Tr}(\rho \log _2{\rho })}$ specifies the von Neumann entropy;
$I_c(\rho _{AB})=\max_{\Pi _A}I[\Pi_A(\rho _{AB})]$
represents the classical correlations of $\rho_{AB}$ and
the maximum in it is taken over all projective measurements $\Pi _A$
on the subsystem $A$.
Other quantifiers of quantum discord include the geometric measure based on the square
norm of trace distance \cite{Dakic}, and distance-based measures either in terms of
the relative entropy \cite{modi} or in terms of the fidelity \cite{brub}.

Generally, the dynamics of the above measures of discord under decoherence will
manifest distinct behaviors from each other quantitatively or even
qualitatively, which suggests the complexity of the
issue of decoherence from different perspectives.
For the species of Bell-diagonal states, it is applaudable that the amount of discord
defined via the information-theoretic measure is
identical to that defined via the distance measure in terms of the relative
entropy \cite{modi}.
At this stage, the freezing and sudden-transition phenomena of discord exhibited
in the Bell-diagonal states are of
particular interest as the dynamical behaviors of both quantum and
classical correlations there are consistent from different views
related to the two kind of measures. We mention that comparing with those
sudden-transition phenomena without exhibiting plateaus \cite{Maziero,Guo,Celeri},
the freezing dynamics of discord requires more critical conditions.
Up to our knowledge, this freezing phenomenon was demonstrated
mostly in systems with initial states
that are of standard Bell-diagonal form with
rank two \cite{Mazzola,MAZZOLA2,experi,feng2,bellomo}.
A complete description of the condition related to the phenomenon is still lacking.

In this paper we investigate necessary and sufficient conditions
of the freezing phenomenon of quantum discord occurring in
the phase damping process. By elaborating the time variation
of the discord of standard Bell-diagonal states, we associate the expression of it
with the function of relative entropy.
This enables us to give a complete description of conditions
for the phenomenon by virtue of the convexity property
of the relative entropy. The overall geometric structure is then depicted
in the parametric space for the collection of states that satisfy these conditions.
It is shown that the derived conditions are applicable also to the
phenomenon occurring in the non-Markovian dephasing process.
Finally, we display that the same condition and phenomenon coincide
in a new sort of Bell-diagonal states beyond the standard form.

\textit{Time variation of quantum discord of Bell-diagonal states under phase damping.}
The phasing damping dynamics we are going to investigate characterizes
the type of quantum noise process that induces loss of quantum coherence
but without energy exchange.
For a two-level system, the dynamical evolution of the state under the phase damping
channel is described as
\begin{equation}
\rho (t)=\left(
\begin{array}{ll}
\rho _{11}(0) & d(t)\rho _{12}(0) \\
d(t)\rho _{21}(0) & \rho _{22}(0)
\end{array}
\right) ,  \label{dephasing}
\end{equation}
where $d(t)$ depicts the degradation of the coherence and it often
takes an exponential form for a Markovian dephasing process.
Hereafter, we shall introduce a notation $q\equiv1-d(t)$ and treat it
a parametrized time.

Let us consider the local phase damping process on a two-qubit system
that initially inhabits in the standard Bell-diagonal state
\begin{equation}
\rho _{AB}=\frac 14(I_4+\sum_{i=1}^3c_i\sigma _i^A\otimes \sigma
_i^B)=\sum_{i=1}^4\lambda _i|\psi _i\rangle \langle \psi _i|,  \label{bell1}
\end{equation}
where $\sigma_i $'s are Pauli operators and $|\psi _i\rangle $'s are standard
Bell bases of the form
$|\psi _{1,3}\rangle =\frac 1{\sqrt{2}}(|00\rangle \pm |11\rangle )$,
$|\psi _{2,4}\rangle =\frac 1{\sqrt{2}}(|01\rangle \pm |10\rangle )$.
The corresponding eigenvalues $\lambda _i\geq 0$ $(i=1,\cdots,4)$
relate to the parameters $c_i$'s via
\begin{equation}
\lambda _{1,3}=\frac 14[1\pm c_1\mp c_2+c_3],~~
\lambda _{2,4}=\frac 14[1\pm c_1\pm c_2-c_3].  \label{relation}
\end{equation}
As one of the subsystems $A$ or $B$ passes through
the channel of Eq. (\ref{dephasing}), the evolving state $\rho _{AB}(q)$
will remain to
be of form of Eq. (\ref{bell1}) and the corresponding eigenvalues
\begin{equation}
\begin{split}
\lambda _{1,3}(q)=\lambda _{1,3} \pm \frac q2 [\lambda _{3}-
\lambda _{1}],  \\
\lambda _{2,4}(q)=\lambda _{2,4} \pm \frac q2 [\lambda _{4}-
\lambda _{2}].
\end{split}
\label{lambda}
\end{equation}
The amount of discord defined via the two kind of measures, the
information-theoretic measure and the distance measure of relative entropy,
coincides for this state and it can be derived analytically
as \cite{Luo}
\begin{equation}
Q_{AB}(q)=1+\sum_{i=1}^4\lambda _i(q)\log _2\lambda _i(q)+h_2(x_M).
\label{discord}
\end{equation}
Here, $h_2(x)$ accounts for a binary entropy function $h_2(x)=-x\log
_2x-(1-x)\log _2(1-x)$, and $x_M(q)=[1+c_M(q)]/2$
in which $c_M(q)\equiv \max \{|c_1(q)|,|c_2(q)|,|c_3|\}$
and $c_{1,2}(q)=(1-q)c_{1,2}$.

The variation of the discord depends on its local analytical property,
recorded as $Q_{AB}^{(i)}(q)$ $(i=1,2,3)$ for the corresponding time intervals
in which $c_M(q)$ equals to $|c_1(q)|$, $|c_2(q)|$, or $|c_3|$,
hence $x_M(q)=\lambda _1(q)+\lambda _2(q)$,
$\lambda _2(q)+\lambda _3(q)$, or $\lambda _1+\lambda _3$, respectively \cite{note}.
Consider first the situation of $c_M(q)=|c_3|$. In this case the term $h_2(x_M)$ in
Eq. (\ref{discord}) becomes a constant, therefore one obtains
\begin{equation}
\frac{\mathrm{d}}{\mathrm{d}q}Q_{AB}^{(3)}(q)=-\frac{\mathrm{d}}{\mathrm{d}q}
S_{AB}(q).
\label{varq3}
\end{equation}
It is direct to observe that for a Markovian channel,
the above variation is always negative since the entropy of $\rho
_{AB}(q)$ should be monotonically increasing under such a dephasing channel.
In order to derive further the variation of $Q_{AB}^{(1,2)}(q)$,
we introduce the notation of relative entropy: $H_n(x\Vert
y)=\sum_{i=1}^nx_i(\log _2x_i/y_i)$, where $\{x_i\}$ and $\{y_i\}$
stand for two $n$-component random variables.
By virtue of the following relation
\begin{eqnarray}
\begin{split}
\frac{\mathrm{d}}{\mathrm{d}q}\lambda _{1,3}(q) &=&\pm \frac
1{2(1-q)}[\lambda _3(q)-\lambda _1(q)], \\
\frac{\mathrm{d}}{\mathrm{d}q}\lambda _{2,4}(q) &=&\pm \frac
1{2(1-q)}[\lambda _4(q)-\lambda _2(q)],
\end{split}
\label{deriv-par}
\end{eqnarray}
we are able to work out that
\begin{equation}
\frac{\mathrm{d}}{\mathrm{d}q}Q_{AB}^{(1)}(q)=\frac 1{2(1-q)}[H_2(\mathbf{
\mu }\Vert \mathbf{\nu })-H_4(\mathbf{\xi }\Vert \mathbf{\eta })],
\label{deriv}
\end{equation}
where $\mathbf{\mu }(\mathbf{\nu })$ and $\mathbf{\xi }(\mathbf{\eta })$
account for probability distributions specified as
\begin{eqnarray}
\begin{split}
\mathbf{\mu }(q) &=\{x_M,1-x_M\},~~\mathbf{\nu }(q)=\{1-x_M,x_M\},
\\
\mathbf{\xi }(q) &=\{\lambda _1(q),\lambda _2(q),\lambda _3(q),\lambda
_4(q)\},  \\
~\mathbf{\eta }(q) &=\{\lambda _3(q),\lambda _4(q),\lambda _1(q),\lambda
_2(q)\}.  \label{random}
\end{split}
\end{eqnarray}
A similar expression can be obtained for $\mathrm{d}Q_{AB}^{(2)}(q)/\mathrm{d}q$
merely through exchanging the roles of $\lambda _1(q)$ and $\lambda _3(q)$
in Eq. (\ref{deriv}).

The explicit expression of the local variation of discord derived above enables
us to analyze its dynamical behavior conveniently.
It is readily seen, from the joint convexity property of
relative entropy \cite{convex1}, that
the terms in Eq. (\ref{deriv}) satisfy
\begin{equation}
H_2(\mathbf{\mu }\Vert \mathbf{\nu })\leq \alpha H_2(\mathbf{\xi }_1\Vert
\mathbf{\eta }_1)+\beta H_2(\mathbf{\xi }_2\Vert \mathbf{\eta }_2)
=H_4(\mathbf{\xi }\Vert \mathbf{\eta }),
\label{jconvex}
\end{equation}
where $\alpha =\lambda _1(q)+\lambda _3(q)$, $\beta =\lambda _2(q)+\lambda
_4(q)$, and $\xi _i$ and $\eta _i$ are variables of probability distributions
given by
\begin{eqnarray}
\begin{split}
\mathbf{\xi }_1(q)&=&\frac 1\alpha \{\lambda _1(q),\lambda _3(q)\},~\mathbf{
\eta }_1(q)=\frac 1\alpha \{\lambda _3(q),\lambda _1(q)\}, \\
\mathbf{\xi }_2(q)&=&\frac 1\beta \{\lambda _2(q),\lambda _4(q)\},~\mathbf{
\eta }_2(q)=\frac 1\beta \{\lambda _4(q),\lambda _2(q)\}.
\end{split}
\label{random4}
\end{eqnarray}
Thus one derives $\mathrm{d}Q_{AB}(q)/\mathrm{d}q\leq 0$
in all time intervals for
the Markovian process. We note that this monotonicity
of quantum discord is a natural consequence since
the Markovian dephasing channel is always unital \cite{brub}.

\textit{Conditions for the freezing and sudden-transtion dynamics of
quantum discord under phase damping}.

\textbf{Theorem}. The Bell-diagonal states of Eq. (\ref{bell1}) undergoing
the phase damping process can exhibit zero quantum discord rate,
i.e., the so-called freezing phenomenon of discord,
if and only if the parameters $\lambda _i$'s satisfy
\begin{subequations}
\begin{align}
\lambda _1\lambda _4 &=\lambda _2\lambda _3,~(\lambda _1-\lambda
_4)(\lambda _2-\lambda _3)>0,
\label{condition1}\\
\mathrm{or}~~~\lambda _1\lambda _2 &=\lambda _3\lambda _4,~(\lambda
_1-\lambda _2)(\lambda _4-\lambda _3)>0,  \label{condition2}
\end{align}
\end{subequations}
alternatively.

We specify that the inequalities in Eqs. (\ref{condition1})
and (\ref{condition2}) figure out
different order relations for $\lambda_i$'s. For instance,
Eq. (\ref{condition1}) indicates the possible order as:
\begin{equation}
\begin{split}
\lambda_1>\lambda_2>\lambda_3>\lambda_4,~~
\lambda_4>\lambda_3>\lambda_2>\lambda_1, \\
\lambda_3>\lambda_4>\lambda_1>\lambda_2,~~
\lambda_2>\lambda_1>\lambda_4>\lambda_3.
\end{split}
\label{order}
\end{equation}

To prove the theorem, let us suppose that one starts from a state of
Eq. (\ref{bell1}) in which the parameters $\lambda_i$'s
satisfy one of the order relations
of Eq. (\ref{order}). Note that any of these order relations yields that $c_M=|c_1|$,
hence $Q_{AB}=Q_{AB}^{(1)}$, for the initial state. Consequently,
the local variation of $Q_{AB}(q)$ during the starting time interval is
illustrated by Eqs. (\ref{deriv})-(\ref{random4}).
From the fact that the relative entropy $H_2(\mathbf{\mu }\Vert \mathbf{\nu })$ is
a strict convex function, the equality in between the first two expressions
of Eq. (\ref{jconvex})
holds if and only if there are
\begin{equation}
\mathbf{\mu }(q)=\mathbf{\xi }_1(q)=\mathbf{\xi }_2(q),~~
\mathbf{\nu }(q)=\mathbf{\eta }_1(q)=\mathbf{\eta }_2(q).
\label{proof}
\end{equation}
This leads clearly to the relation $\lambda _1\lambda _4=\lambda _2\lambda_3$,
which is responsible for $\mathrm{d}Q_{AB}(0)/\mathrm{d}t=0$ of
the specified initial state.
It is crucial to note that according to Eq. (\ref{lambda}),
the relation $\lambda_1(q)\lambda_4(q)=\lambda_2(q)\lambda_3(q)$
will sustain during the evolution. The zero-rate dynamics of
discord thus will continue until the
presumed order relation is broken. In addition,
it is direct to observe that a similar discussion is
applicable for the condition of Eq. (\ref{condition2}) if
one starts from a state with $c_M=|c_2|$. These facts together with that
$\mathrm{d}Q_{AB}^{(3)}(q)/\mathrm{d}q\neq 0$ [cf. Eq. (\ref{varq3})]
then complete our proof for the theorem.
Straightforwardly, by substituting Eqs. (\ref{condition1})
or (\ref{condition2}) into Eq. (\ref{discord}) one obtains that
\begin{equation}
Q_{AB}(q)=1-h_2(\lambda _1+\lambda _3)=Q_{AB}(0),
\label{traj1}
\end{equation}
where $h_2(x)$ stands for the aforementioned binary entropy function.

To shed light further on the physics related to the above derived condition of
the freezing phenomenon, we notice that the term $H_2(\mu (q)\Vert \nu (q))$
in Eq. (\ref{deriv}) can be rewritten as
$H_2(\mu (q)\Vert \nu (q))=H_4(\xi ^{\prime }(q)\Vert \eta ^{\prime}(q))$,
where
\begin{equation}
\begin{split}
\xi ^{\prime }(q) =\{\frac{x_M(q)}2,
\frac{x_M(q)}2,\frac{1-x_M(q)}2,
\frac{1-x_M(q)}2\},
\\
\eta ^{\prime }(q) =\{\frac{1-x_M(q)}2,
\frac{1-x_M(q)}2,\frac{x_M(q)}2,
\frac{x_M(q)}2\},
\end{split}
\label{mixing}
\end{equation}
and $x_M(q)=\lambda_1(q)+\lambda_2(q)$. Recall that the relative entropy
$H_n(x\|y)$ tells actually how difficult it is to
distinguish two states of the stochastic event, $\{x_i\}$ and
$\{y_i\}$. The right hand of Eq. (\ref{deriv}) can be understood
as the reduction of the distinguishability of the
two states undergoing a mixture process: $\xi(q)\rightarrow \xi^{\prime }(q)$
and $\eta(q)\rightarrow \eta^{\prime}(q)$, that is, the components of the first two
of $\xi(q)$ and $\eta(q)$ are mixed and so well as the last two, respectively.
At this stage, an alternative perspective on the freezing phenomenon
of discord, respecting its necessary and sufficient conditions (\ref{condition1})
and (\ref{condition2}), is that the distinguishability of the two states
$\xi(q)$ and $\eta(q)$ undergoing the specified stochastic mixture process
will not alter as long as the corresponding condition is fulfilled.

With the necessary and sufficient conditions described above, we
are now able to depict in Fig. 1 an overall geometric profile for those
states in terms of parameters
$\{\sqrt{\lambda_1},\sqrt{\lambda_2},\sqrt{\lambda_3}\}$.
Regarding that $0\leq\sum_{i=1}^3\lambda _i\leq1$, the set of standard
Bell-diagonal states constitute the portion of a unit sphere located in the
first quadrant. The surface responsible for the conditions in Eqs. (\ref
{condition1}) and (\ref{condition2}) comprises four leaf-shaped
curves and the only common node of
them corresponds to the maximally random state $\rho_{AB}=I/4$.
The boundary of the surface figures out the collection of transition points
which can be described explicitly by four independent
sets of equations as
\begin{eqnarray}
\begin{split}
\lambda_1&=\lambda_2=\sqrt{\lambda_3}-\lambda_3,
\\
\lambda_3&=\lambda_2=\sqrt{\lambda_1}-\lambda_1,
\\
\lambda_1&=1-\sqrt{\lambda_2}-\lambda_3=
(1-\sqrt{\lambda_2})^2,
\\
\lambda_3&=1-\sqrt{\lambda_2}-\lambda_1=(1-\sqrt{\lambda_2})^2.
\end{split}
\label{edge}
\end{eqnarray}

\begin{figure}[h]
\includegraphics[width=5.4cm]{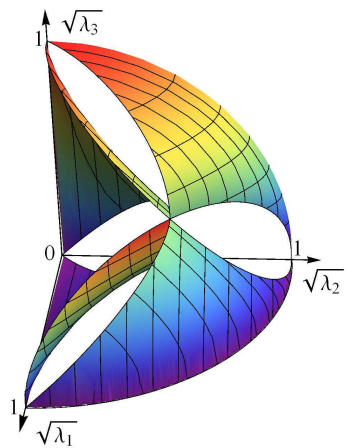}
\caption{Geometric profile in the parametric space of standard Bell-diagonal states
that could exhibit freezing dynamics of quantum discord under phase damping.}
\label{fig1}
\end{figure}

The freezing and sudden transition phenomena of the dynamics are displayed
schematically in Fig. 2.
The trajectory of an initial state at the point $I$ proceeds along the surface with
a constant discord till it encounters the boundary at the point $T$, after which
the discord decreases to zero and the trajectory reaches finally
the point $F$. At the transition point $q=q_T$, the rate of quantum discord
suffers a sudden change from zero to a value of
\begin{eqnarray}
\frac{\mathrm{d}Q_{AB}^{(3)}}{\mathrm{d}q}|_{q=q_T^+} =-\frac{1}{2(1-q_T)}
H_4(\xi(q_T)\Vert\eta(q_T)),
\end{eqnarray}
where we have used the relation of Eq. (\ref{varq3}) and $\xi$ and $\eta$ are
specified in Eq. (\ref{random}).

It is worthwhile to point out that the conditions
described in Eqs. (\ref{condition1}) and (\ref{condition2})
are applicable also to the frozen discord
occurring in the non-Markovian dephasing process \cite{MAZZOLA2}.
The dynamical evolution of this process can still be described
phenomenologically by Eq. (\ref{dephasing}), hence the dependency of the discord on
the parameter $q$ does not alter. As a result, all discussions related to
Eqs. (\ref{condition1}) and (\ref{condition2}) are valid for its dynamics.
Note that in the non-Markovian process $d(t)$ (hence $q$) is not a
monotonic function of $t$ any more. Say, for the channel with random
telegraph signal noise
it is described as \cite{noise}
\begin{equation}
q=1-d(t)=1-e^{-\gamma t}
[\cos(\omega t)+\frac{\gamma}{\omega }\sin(\omega t)],
\label{nonmark}
\end{equation}
where $\omega=\sqrt{4 a^2-\gamma^2}$, and $a$ characterizes the
coupling strength of the system with environment. Differing from the model concerned
in Ref. \onlinecite{MAZZOLA2}, in our system only one of the qubits is subjected
to the noise channel. To display multiple transitions and freezing phenomena
of the discord, the parameters of the initial state herein should satisfy
$|c_{1(2)}/c_3|>e^{\pi \gamma/\omega}$.
In the parametric space of $\{\sqrt{\lambda_1},\sqrt{\lambda_2},\sqrt{\lambda_3}\}$,
we depict the evolution of the non-Markovian
process with $c_1/c_3=\frac {35} {24}$ and
$\gamma/\omega=\frac 1 {32}$.
It is shown that the trajectory manifests a traversing phenomenon
between two separate leafs (see Fig. 2).

\begin{figure}[h]
\includegraphics[width=6.0cm]{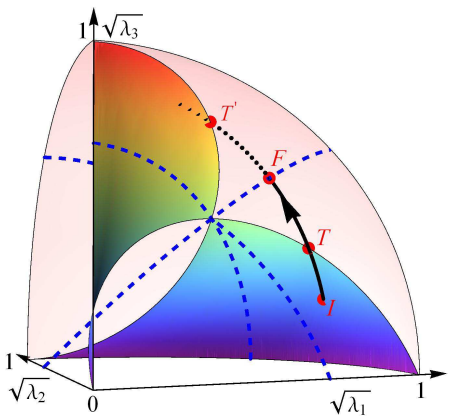}
\caption{Illustration of the trajectory for the freezing
and sudden transition dynamics of quantum discord in the parametric space.
The three blue-dashed lines account for the collection of Bell-diagonal states
with zero discord and they share a common node with the two leaf-shaped curves
determined by Eq. (\ref{condition1}). The initial state is specified as
$\lambda_1=3/4,\lambda_2=3/16,\lambda_3=1/20$.
The trajectory of the solid line proceeding along
$I\rightarrow T\rightarrow F$ describes the evolution of the Markovian
phase damping process.
For a non-Markovian dephasing process specified by Eq. (\ref{nonmark}),
the trajectory of the evolution proceeds further along the dotted line
$F\rightarrow T'$ and oscillates damply between the two specified leafs,
displaying feedback and multiple transitions of the discord dynamics.}
\label{fig2}
\end{figure}

As a final proposal of this paper, we demonstrate that the above described
freezing dynamics of discord and associated conditions will occur and coincide
in a sort of new Bell-diagonal states beyond
the standard form of Eq. (\ref{bell1}). Consider that
\begin{equation}
\rho _{AB}=\frac 14(I_4+\sum_{i=1}^3c_{ij}\sigma _i^A\otimes \sigma
_j^B)=\sum_{i=1}^4\lambda^\prime _i|\psi^\prime _i\rangle \langle \psi^\prime _i|,  \label{bell2}
\end{equation}
where $c_{12}c_{21}\neq0$ but all other off-diagonal elements of $\{c_{ij}\}$ vanish.
The eigenstates are shown as
$|\psi^\prime _{1,3}\rangle =\frac 1{\sqrt{2}}(|00\rangle \pm e^{-i\phi_+}|11\rangle)$,
$|\psi^\prime _{2,4}\rangle =\frac 1{\sqrt{2}}(|01\rangle \pm e^{-i\phi_-}|10\rangle)$
with $\phi_{\pm}=-\arctan \frac {c_{12}\pm c_{21}}{c_{22}\mp c_{11}}$, and the
corresponding eigenvalues $\lambda^\prime _i$'s relate to $c_{ij}$'s via
\begin{equation}
\begin{split}
\lambda ^\prime _{1,3}&=\frac 14[1\pm
\sqrt{(c_{11}-c_{22})^2+(c_{12}+c_{21})^2}+c_{33}], \\
\lambda ^\prime _{2,4}&=\frac 14[1\pm
\sqrt{(c_{11}+c_{22})^2+(c_{12}-c_{21})^2}-c_{33}].
\end{split}
\end{equation}
Suppose that the subsystem $A$ is subjected to the local phase damping channel
described by Eq. (\ref{dephasing}).
It turns out that the state $\rho_{AB}(q)$ will keep the form of
Eq. (\ref{bell2}) with $\{c_{ij}\}$
evolving as: $c_{33}(q)=c_{33}$, and
$c_{ij}(q)=(1-q)c_{ij}$ for all other elements. Consequently,
the eigenvalues of $\rho_{AB}(q)$, $\lambda ^\prime _i(q)$,
are derived promptly and they are verified
to possess the similar expressions as those in Eq. (\ref{lambda}).
This implies that the discord of $\rho_{AB}(q)$ should display the same dynamics as
states of Eq. (\ref{bell1}). Therefore, it is safe to conclude that the
freezing phenomenon could occur for the present state and the conditions of it are
coincident with those in Eqs. (\ref{condition1}) and (\ref{condition2}).

In summary, we have studied freezing phenomena of quantum discord for
Bell-diagonal states under phase damping noise processes.
We have obtained necessary and sufficient conditions for these
intriguing phenomena and the overall geometric profile is then
illustrated in the parametric space for the corresponding states.
Moreover, we have shown that the same freezing dynamics
and conditions could occur and coincide in a sort of new Bell-diagonal
states beyond the standard form of it. For further applications,
our explicit derivation of the condition for frozen discord is of potential
usefulness to realize quantum information processing, e.g.,
designing quantum systems to preserve at best the quantum coherence
in the presence of the noisy environment.

This work was supported by the NSFC under grant No. 10874254.


\begin{thebibliography}{99}
\bibitem{note} Correspondingly, it is noted that during the time interval with
$c_M(q)=|c_i(q)|$, the optimal projective measurement that achieves the maximization
of the classical correlation $I_c(\rho_{AB}(q))$ is given by
$\Pi_A^{\pm}(q)=\frac 12(I_2\pm \sigma _i)$.
\end{thebibliography}
\end{document}